\documentclass{epl}
\usepackage{graphicx}

\begin{document}

\shorttitle{Scaling out}

\title{Scaling out the density dependence of the $\alpha$ relaxation in
glassforming polymers}
\author{C. Alba-Simionesco\inst{1}, A. Cailliaux\inst{1}, A. Alegr\`{\i}a\inst{2},
G.   Tarjus\inst{3}}
\institute
{\inst{1} Laboratoire de Chimie Physique, B\^atiment 349,
Universit\'e de Paris Sud, F-91405 Orsay, France \\
\inst{2} Departamento de F\'\i\'{\i}sica de Materiales,
Facultad de Qu\'\i mica, Universidad del Pa\'\i s Vasco (UPV/EHU),
 Apartado 1072, 20080 San Sebasti\'an, Spain\\
\inst{3}  Laboratoire  de   Physique  Th{\'e}orique des
Liquides,  Universit{\'e} Pierre et  Marie Curie,  4 Place Jussieu,75252
Paris Cedex 05, France }

\date{submitted to Europhys. Lett., March 11, 2004}

\maketitle

\begin{abstract}

We  show  that the  density  and     temperature dependences of  the
$\alpha$-relaxation time of  several glassforming polymers can be described
through a  single  scaling variable $X=e(\rho)/T$,  where  $e(\rho)$ is well
fitted by    a  power law    $\rho^x$,   $x$ being  a  species-specific
parameter.  This  implies   that   ``fragility''  is  an    intrinsic,
density-independent   property  of  a glassformer  characterizing its  super-Arrhenius
slowing down   of relaxations,  and it   leads us to
propose a modification of the celebrated Angell plot.

\end{abstract}


\def\be{\begin{equation}}
\def\ee{\end{equation}}
\def\bea{\begin{eqnarray}}
\def\eea{\end{eqnarray}}
\def\bit{\begin{itemize}}
\def\eit{\end{itemize}}

The   glass transition  of  liquids  and  polymers is  conventionnally
studied at   constant, usually atmospheric,   pressure $(P)$ by cooling the
system.  However, in an effort  to disentangle the effects of  density $(\rho )$
and  temperature $(T)$ that both  influence the  viscous slowing down  under
isobaric conditions and  to provide a  more stringent test of existing
models and theories,  there has recently  been an increasing number of
sytematic studies   of  glass formation  using   both $T$ and $P$
 as experimentally controlled variables. Glassforming polymers
are  interesting  in    this respect : first, because   of   their  rather high
compressibility allowing to cover a significant range of densities and second, because
of  the important  role played  in  polymer science by the free-volume
theory that   puts  the  emphasis on congestion    effects due  to the
increase  of density as one  approaches the glass transition\cite{ferry}. 

In this  letter we analyze new dielectric  relaxation data obtained on
several  glassforming polymers   up  to 3 kbars   poly(epichlorhydrine)
(PECH),  poly(vinylmethylether)  (PVME),  poly(vinylacetate) (PVAc),
poly(methylmethacrylate)  (PMMA)),  and a combination of dielectric, neutron   scattering   and calorimetric  data on
1,4-poly(butadiene) (1,4-PB). We
show that the  $\rho$ and $T$ dependences of the
$\alpha$-relaxation time ($\tau_\alpha$ expressed, say, in seconds) can   be  described by a single    scaling
variable $X=e(\rho)/T$,
\begin{equation}
\label{eq1:equation}
log(\tau_\alpha(\rho,T)) = F(\frac{e(\rho)}{T}), 
\end{equation}	

where, over the range of  densities experimentally accessible,  $e(\rho)$
can be  fitted  by  a power law,  $\rho^x$;  the  parameter  $x$  and the
function $F(X)$ are  species-specific.  This  scaling generalizes   to
polymers   work recently  done  on  several  molecular  liquids.  As a
spin-off, we  show that ``fragility'', a  concept introduced by Angell\cite{angell85}
to characterize the degree of deviation from Arrhenius behavior in the
$T$-dependence of $log(\tau_\alpha)$  as one approaches  the glass transition,
is independent of density  and is  thus  an inherent property of  each
glassformer. This leads us  to suggest  a  modification of  the widely
used Angell plot.

As a   more extensive  data   base  on the  $P$ and   $T$
dependences   of  the $\alpha$-relaxation  time (and  of  the viscosity)  of
glassforming  polymers  and  liquids now becomes  available\cite{schug98},   and
reasonably accurate equations of state  exist to convert $P,T$ results
to  $\rho,T$ results,  the question arises  of how  to  best organize the
data, in a model-free and  physically meaningful manner.  In the  case
of molecular liquids, it has  been recently suggested that the $\rho $
dependence could be described  through a single parameter, an  effective
activation energy  $E_{\infty }(\rho)$ characteristic  of  the  high-$T$
Arrhenius regime,  i.e., $log(\tau_\alpha(\rho,T))  =  F(E_{\infty }(\rho)/T)$.  The
same procedure, however, is not   applicable to glassforming  polymers
because most of them cannot be studied at $T$ high enough for
reaching  an   Arrhenius    regime,    so that
$E_{\infty }(\rho)$  cannot be directly determined.   In the cases  studied in
\cite{CAS02} $E_{\infty }(\rho)$ was a  quite featureless, increasing function
of $\rho$ (in the accessible  range); rather than using an (uncontrolled)
extrapolation of the data at high $T$, it  then seems more sensible to
choose a   simple predefined  function  of  $\rho$, $e(\rho)$,  with  as few
adjustable   parameters   as possible, to    build a  scaling variable
$X=e(\rho)/T$.  An appealing choice is a  one-parameter power law, $\rho^x$,
$x$ being material-specific. Such a choice has  been used to represent
$\alpha$-relaxation  time data   for    the fragile glassforming     liquid
ortho-terphenyl   (o-TP), with    $x=4$\cite{toelle01};    it is  also
consistent with  a   recent study of   the  density dependence of  the
glass-transition line $T_g(\rho)$ of  several polymers (up to few kbars),
where it   was  found that  $T_g(\rho)\alpha\rho^x$,    with $x\approx2.0$ for  atactic
poly(propylene), $x\approx  2.7$     for poly(styrene), and    $x\approx 3.4$  for
poly(carbonate)\cite{hollander01}.     We  have   thus  analyzed   the
$\alpha$-relaxation time $\tau_\alpha(\rho,T)$,  obtained  from dielectric  relaxation
measurements made on  PECH, PVME, PVAc, and PMMA  over a wide range of
$T$  and  $P$ (all  experimental   details are given  in \cite{aude03,
chauty03, Alegria}) and converted to $(\rho,T)$ via the equation of state
proposed by Sanchez and Cho \cite{PVT}, as a  function of $\rho^x/T$, $x$
being an adjustable parameter. As illustrated in Fig. 1
, a very good scaling of the
data is  indeed reached.  We  find $x\approx 2.7$ for PECH   and PVME, $x\approx 1.4$ for
PVAc, and $x\approx 1.25$  for PMMA.  We have  applied  the same analysis to  the
$\alpha$-relaxation time  data  obtained  on 1,4-PB \cite{14PB} with $x\approx 1.8$. In
all cases a  good data collapse is  found with Eq. (1)
and $e(\rho)=\rho^x/T$ :  see inset of Fig. 4.

\begin{figure}[t]
\centering
\resizebox{10cm}{!}{\includegraphics [height=6cm]{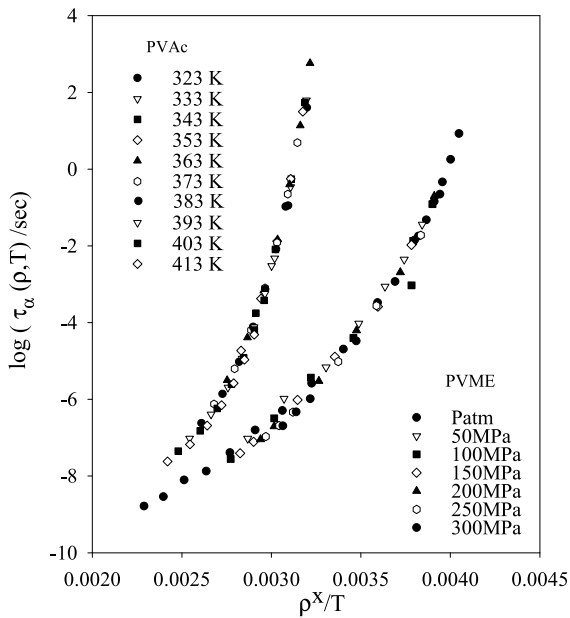}}
\caption{
Scaled plot of
$log(\tau _\alpha (\rho ,T))$ 
versus $X = \rho ^x/T$ for PVME($x=2.7$) and
PVAc($x=1.4$). Data were obtained at atmospheric pressure and along different isotherms and
 isobars up to 300MPa.}
\label{fig1}
\end{figure}

Before discussing some of the consequences of the scaling, it is worth
stressing that despite its appealing  form, the power law behavior  of
$e(\rho)$ may not convey much physical content. For o-TP it was suggested
that   the $\rho^4/T$ scaling was   reminiscent of the well-known scaling
property of  simple  liquids  interacting  via  a soft repulsive  pair
potential,   $v(r)=\epsilon(\frac{\sigma}{r})^n$,   with   $n=12$   as    in   the
Lennard-Jones    model\cite{Hansen}:   indeed, for   such  systems the
long-time  properties  depend on  $\rho $ and   $T$ only through a single
combination  $\rho  T^{-3/n}$,  hence   for   $n=12$,  $\rho  T^{-1/4}$,  or
equivalently, $\rho^4/T$. However,   such  an interpretation  implies   a
strong corollary, that    the excess thermodynamic  properties of  the
system (measured  with respect to the   ideal gas contribution) should
also  depend on the   scaling  variable, $\rho T^{-3/n}$  or $\rho^{n/3}/T$,
only. We  have checked  that this  is  not the   case for any   of the
polymers considered here (nor for o-TP): as  illustrated in Fig. 2 for
PVME and PB, the quantity $(\frac{P}{\rho T})$ obtained from the existing
equations of  state  is not  a  unique  function of $\rho^x/T$  with  $x$
determined  from the scaling  plot   of the $\alpha$-relaxation time.   Not
unexpectedly, in the available range of $\rho  $,
 polymers do not behave
as  soft  repulsive spheres.   The   power law $\rho^x$ should    thus be
considered as a convenient, one-parameter way of representing the $\rho $
dependence of the activation energy scale $\propto e(\rho)$, a scale that, when
data  are available  at high enough  $T$, can  be  identified with the
empirical     Arrhenius    activation    energy      $E_\infty    (\rho     )$
\cite{CAS02}. Actually, a good agreement  to the $\alpha $-relaxation  time
data  can  also  be obtained    by choosing   a one-parameter   linear
description, $e(\rho)=\rho -\rho ^*$  : see Fig. 3.  Again, no special emphasis
should be put  on the fact  that $e(\rho)$ is a linear  function of $\rho $;
this merely reflects the  fact that over the  limited range of density
that is accessible experimentally, $e(\rho)$ increases monotonically with
$\rho$ in a featureless fashion.

\begin{figure}[t]
\resizebox{12cm}{!}{\includegraphics{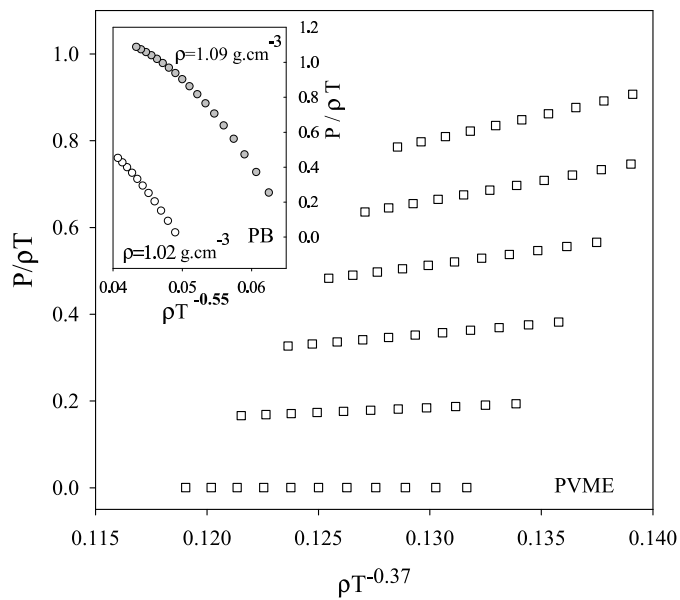}}%
\caption{
Equation of state $P/\rho T$ versus $X=\rho ^x/T$ for PVME calculated from \cite{PVT} along isobars; 
inset : 1,4-PB along two isochores.
Note the absence of scaling.}
\label{fig2}
\end{figure}

Besides  the fact  that  it  helps  organizing  all experimental  data
obtained by changing $P$ and/or $T$ in a simple  and rational way, the
main consequence of the  scaling expressed by Eq.~(\ref{eq1:equation})
is   that  "fragility",   which  characterizes   the   super-Arrhenius
$T$-dependence of the  $\tau _{\alpha } $  and the viscosity,  is an intrinsic
property of a glassformer, in that it does not  vary with density. The
concept of fragility has  proved  to be  most  useful in the  study of
glassforming liquids and polymers and is  now part of the very lexicon
of such studies. Quantifying  the degree of fragility  of a system can
be  conveniently done without having  recourse  to fitting formulae by
considering the steepness  index at  constant  density,  $m_{\rho ,\tau  }$,
defined as
\begin{figure}[t]
\centering
\resizebox{10cm}{!}{\includegraphics{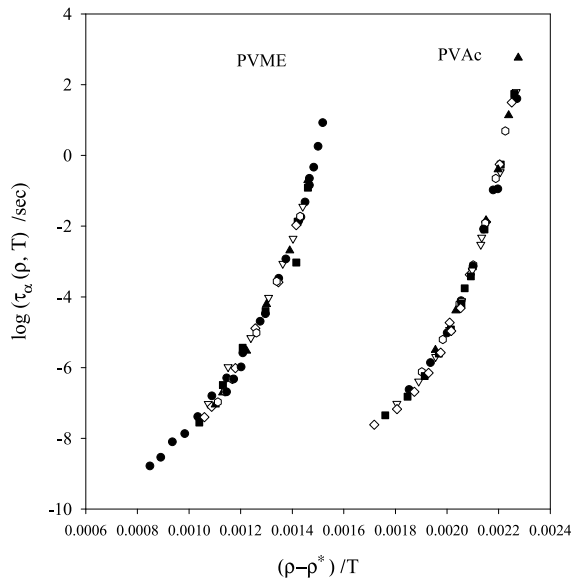}}
\caption{
Scaled plot $log(\tau _\alpha (\rho ,T) )$ versus $X=e(\rho )/T$ with $e(\rho )=
\rho -\rho ^*$ for PVME ($\rho ^*=0.62$) and PVAc ($\rho ^*=0.29$); same symbols as in Fig.1.}
\label{fig3}
\end{figure}

\begin{equation}
\label{eq2:equation}
m_{\rho ,\tau }= \frac{\partial log(\tau_\alpha)}{\partial (\frac{T_\tau }{T})}\bigg|_{\rho }(T=T_\tau ), 
\end{equation}	
where $T_\tau $ is the temperature at  which $\tau _\alpha  =\tau $ (expressed, say,
in  sec.)  at the  given $\rho  $.  By using Eq.~(\ref{eq1:equation}) and
defining $X_\tau $ as the value of  the scaling variable such that $F(X_\tau
)= log(\tau )$, it is  easy to show that  $m_{\rho ,\tau }=X_\tau F'(X_\tau )$, where
$F'(X_\tau  )=\frac{dF(X)}{dX}$,  is {\em  independent  of density}. Thus
fragility,  when properly defined,  is  an intrinsic  property  of the
glassformer. Breakdown of this feature may occur when major structural
changes take place in the system under compression, as for instance in
tetrahedrally  bonded liquids such as $SiO_2$  or  $H_2O$ in which the
local  coordination is known  to   change with density. The  empirical
observation  that   fragility may  vary  with   {\em  pressure}  is  a
consequence of the  possible change in  the relative  contributions to
slowing down due to $T$  and to $\rho $.   Indeed, the steepness index at
constant pressure, $m_{P,\tau }$, defined    by an equation similar    to
Eq.~(\ref{eq2:equation})  is   related to   that  at  constant density
through

\begin{figure}[t]
\centering
\resizebox{10cm}{!}{\includegraphics{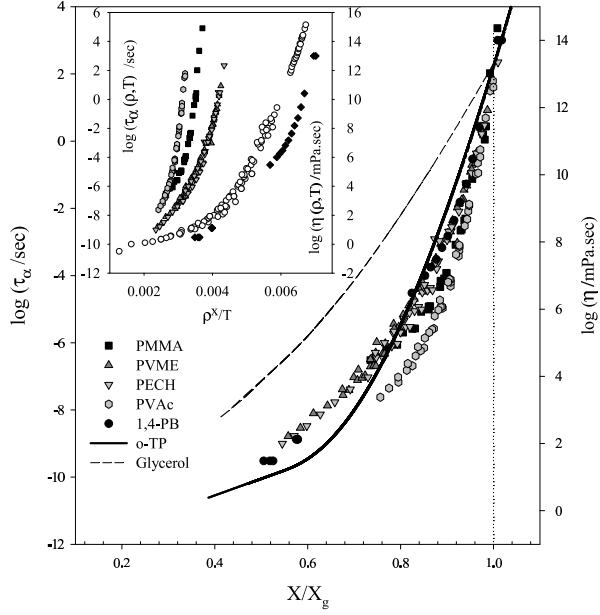}}
\caption{
Modified Angell plot : $log(\tau _\alpha )$ versus  $X/X_g$ where $X_g$ is the
glass transition value  ($\tau _\alpha = 100  sec$ or $\eta =  10^{13}mPa.s$) for
all polymers shown in inset.  For comparison we  also show as full and
dashed   lines   the  scaled curves for     o-terphenyl  and glycerol,
respectively\cite{CAS02}.   Inset : scaled  plot  $log(\tau  _\alpha (\rho ,  T)$
versus  $X=\rho   ^x/T$   for  several polymers,    PVME,  PECH($x=2.7$),
PMMA($x=1.25$),   PVAc($x=1.4$),   1,4-PB   ($x=1.8$)  and o-terphenyl
($x=4$).}
\label{fig4}
\end{figure}

\begin{equation}
\label{eq3:equation}
m_{P ,\tau }(P)= m_{\rho ,\tau }(1+\alpha_P/|\alpha_\tau|),
\end{equation}	

where $\alpha_P=-\rho^{-1}(\partial \rho/\partial T)_P$   and $\alpha_\tau=-\rho^{-1}(\partial \rho/\partial T)_\tau$ are  the
isobaric and   isochronic  coefficients  of  expansivity; the    ratio
$\alpha_P/|\alpha_\tau|$ characterizes the relative effect of $\rho $ over that of $T$
at constant $P$ when  $\tau  _\alpha =\tau $\cite{ferrer98}.  $(1+\alpha_P/|\alpha_\tau|)$  is
also   equal   to           $\frac{H_P}{E_V}$\cite{naoki87},      with
$H_P=(\partial\ln(\tau_\alpha)/\partial(1/T))_P$ and $E_V=(\partial\ln(\tau_\alpha)/\partial(1/T))_\rho$).  One   can
see   from Eq.~(\ref{eq3:equation})   that  the fragility  measure  at
constant   $P$ is always larger  than  the constant-$\rho $ fragility and
that its variation with $P$ depends  on that of the ratio $\alpha_P/|\alpha_\tau|$;
the  amplitude and the sign  of this latter  is species-specific : for
instance it has been shown to decrease with increasing $P$ in the case
of  glycerol  (resulting   in  a decrease    of  the  fragility   with
pressure\cite{cook94})  and  to  increase  with $P$    in  the case of
1,4-polybutadiene\cite{ferrer98}.  Combining  Eq.~(\ref{eq1:equation})
and 
 Eq.~(\ref{eq3:equation}) gives
%
$\alpha_P/|\alpha_\tau| = \alpha _P \frac{dln (e(\rho ))}{dln(\rho) } T_\tau$,
%
%
  which in the case of a power law behavior $e(\rho )=\rho ^x$ simply reduces to $\alpha _P. x .T_\tau $, where both
$\alpha _P$ and $T_\tau $ depend on $P$; usually $\alpha _P$ decreases with $P$\cite{alba91} and $T_\tau $ increases
 with $P$, lead to a variety of behavior for $\alpha_P/|\alpha_\tau|$ (see above).

The fact that  fragility is an intrinsic  property of a glassformer is
best  represented     by     modifying  the    now    standard  Angell
plot\cite{angell85} in which $log(\tau _\alpha  )$ is shown versus the inverse
scaled temperature  $T_g/T$ at constant (usually atmospheric) pressure
: we suggest  instead, when enough  data is available,  to plot $log(\tau
_\alpha) $  versus $X/X_g$,  where $X= e(\rho  )/T$   is the scaling  variable
introduced  above  and  $X_g$  its  value  when   $\tau _\alpha $    reaches a
characteristic "glass transition"  value,  say  $\tau  _\alpha =  100sec$  for
dielectric  relaxation    data.  The  steepness   of  the   $log(\tau  _\alpha
)$-vs-$X/X_g$  curve is  a  measure of the   intrinsic fragility  of a
system, since indeed, according to Eq.~(\ref{eq1:equation}),

\begin{equation}
\label{eq4:equation}
m_X= \frac{dlog(\tau _\alpha)}{d(X/X_g)}\bigg|_{X=X_g}= X_g F'(X_g),
\end{equation}	

is equal to $m_{\rho   ,\tau }$ at the   chosen glass transition  point. The
modified Angell plot is  shown in  Fig. 4  where all the  polymer data
considered in this study (see inset  of Fig. 4) are displayed together
with results for a "fragile"  liquid o-terphenyl and an "intermediate"
one, glycerol\cite{CAS02}.

In conclusion, we have shown  that the $P$  and $T$ dependences of the
$\alpha$-relaxation  time of glassforming  polymers    and liquids can   be
combined, after conversion to $\rho - T$ data, in a  function of a single
scaling variable $X =e(\rho)/T$, where $e(\rho )$ is well  fitted by a power
law $\rho  ^x$, $x$ being species-specific.  On  the practical side, this
provides a simple way to organize and display the increasing amount of
isobaric and isothermal  relaxation time data on glassforming polymers
and liquids. On the theoretical side,  this suggests that fragility is
an intrinsic property of a glassformer, which is best illustrated on a
modified Angell  plot  $log(\tau _\alpha )$  versus $X/X_g$,  and that $e(\rho )$
characterizes the density dependence  of the "bare"  activation energy
scale.

\section*{Acknowledgments} This work has been funded by the CNRS, University Paris-Sud,
University P.   et M.  Curie,  in  spite of the   lack of  support  to
fundamental research from the present French Government.

 We would like to dedicate this letter to our late coworker and friend
 Daniel Kivelson whose  stimulating presence and creative thinking  we
 miss very much.

\end{document}